\documentclass[prl,twocolumn,superscriptaddress,showpacs,amsmath,amssymb,aps]{revtex4-1}
\usepackage{amsmath}
\usepackage{amssymb}
\usepackage{graphics}

\usepackage{graphicx}
\usepackage{dcolumn}
\usepackage{txfonts}
\usepackage[pdfpagemode=UseNone,pdfstartview=FitH,colorlinks=true,linkcolor=blue,
urlcolor=blue,anchorcolor=blue,citecolor=blue]{hyperref}
\allowdisplaybreaks[4]

\usepackage{color}
\usepackage{soul,xcolor}
\usepackage{ulem}
\setstcolor{red}

\usepackage[columnwise]{lineno}

\begin{document}

\title{Topological phase transition  in  a narrow bandgap semiconductor nanolayer}
\author{Zhi-Hai Liu }
\affiliation{Beijing Academy of Quantum Information Sciences, Beijing 100193, China}
\affiliation{Beijing Key Laboratory of Quantum Devices, Key Laboratory for the Physics and Chemistry of Nanodevices, and School of Electronics, Peking University, Beijing 100871, China}

\author{Wenkai Lou}
\affiliation{SKLSM, Institute of Semiconductors, Chinese Academy of Sciences, Beijing 100083,
 China}

\author{Kai Chang}
\affiliation{SKLSM, Institute of Semiconductors, Chinese Academy of Sciences, Beijing 100083,
 China}

\author{H.~Q. Xu}
\email{hqxu@pku.edu.cn}
\affiliation{Beijing Academy of Quantum Information Sciences, Beijing 100193, China}
\affiliation{Beijing Key Laboratory of Quantum Devices, Key Laboratory for the Physics and Chemistry of Nanodevices, and School of Electronics, Peking University, Beijing 100871, China}

\begin{abstract}
Narrow bandgap semiconductor nanostructures have been explored for realization of topological superconducting quantum devices in which Majorana states can be created and employed for constructing topological qubits. However, a prerequisite to achieve the topological phase transition in these nanostructures is application of a magnetic field, which could complicate the technology development towards topological quantum computing. Here we demonstrate that a topological  phase transition can be achieved in a narrow bandgap semiconductor nanolayer under application of a perpendicular electric field. Based on  full band structure calculations, it is shown that the  topological phase transition occurs at an electric-field induced band inversion and is accompanied by a sharp change of the $\mathbb{Z}_{2}$ invariant at the critical field. We also demonstrate  that the nontrivial topological phase  is manifested by the quantum spin Hall edge states in a band-inverted  nanolayer Hall-bar structure.  We present the phase diagram of the nanolayer in the space of  layer thickness and electric field strength,  and discuss  the  optimal conditions  to achieve a large topological bandgap in the electric-field induced  topological phase of a semiconductor nanolayer.  
\end{abstract}
\date{\today}
\maketitle

{\it Introduction}.---Recently, narrow bandgap semiconductor nanostructures have attracted intensive attentions for exploring new topological phases of matter~\cite{Qi2011,Frolov2020,Hasan2010,Lutchym2018}. Free-standing narrow bandgap semiconductor InAs and InSb nanowires can be made in proximity to $s$-wave superconductors to create zero-energy Majorana bound states (MBSs) due to the intrinsic strong spin-orbit couplings (SOCs) in the materials~\cite{Lutchyn2010,Oreg2010}. Although the signatures of single MBSs have been detected in semiconductor-superconductor heterostructured nanowires in recent years~\cite{Mourik2012, Deng2012, Deng2016, Aghaee2023}, it is still experimentally challenging  to integrate  two or more  pairs of MBSs together in these one-dimensional (1D) systems for implementing topological quantum computing~\cite{Alicea2011,Aasen2016,Aguado2020,Sarma2015}. By assuming the coexistence of superconductivity and topological order in a two-dimensional (2D) topological insulator (TI), it is theoretically predicted that the creation and manipulation of topological qubits can be realized~\cite{Fu2008,Schulz2020,Heck2015,Guo2023,Lian2018}.
However,  conventional  narrow bandgap semiconductor layers, such as free-standing InAs and InSb nanolayers~\cite{Mata2016,Pan2016,Pan2019}, with good superconducting proximity effects and strong SOCs, have usually but widely exploited  in their trivial  phase \cite{Zhi2019,Kang2019,Chen2021,Zhang2022,Fan2022,Zhang2022-2,Yan2023}. Thus, it is  fundamentally important to show whether a nontrivial topological phase can be induced into these conventional layered semiconductor materials using an experimentally accessible, controlled method.

\begin{figure}[b]
\includegraphics[width=0.48
\textwidth]{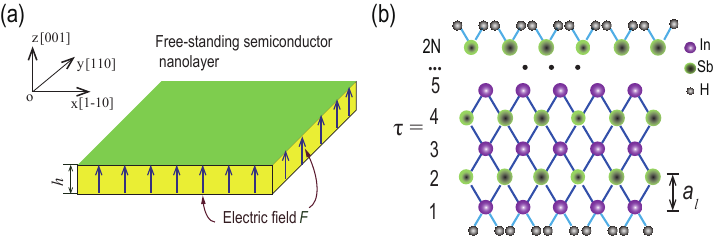}
\caption{(a) Schematics for a freestanding  InSb nanolayer  in the presence of an perpendicular electric field $F$ along the thickness [001] direction.  (b) Atomic structure of an InSb nanolayer, where $\tau=1,2,\cdots,2N$ are the indices for atomic layers, $a^{}_{l}$  is the inter-layer distance, and $2N$ is the number of atomic layers in the nanolayer.}
\label{Fig1}
\end{figure}

Here, in this article, we propose a  new but experimentally convenient way of introducing a 2D TI phase in a planar narrow bandgap semiconductor nanolayer. We will demonstrate that a perpendicular electrical field can be employed to manipulate the energy bands and to induce  a topological phase transition in the semiconductor nanolayer. Experimentally, application of such a perpendicular electrical field to a semiconductor nanolayer can be achieved via a planar dual-gate technique \cite{Chen2021,Zhang2022,Fan2022}. The demonstration is based on solid calculations for the full band structures and the $\mathbb{Z}_{2}$ invariant of a typical semiconductor, such as an InSb nanolayer,  under different strengths of the applied electrical field. We will show that the energy bands will be inverted, when the electrical field becomes larger than a critical strength, and the nanolayer will undergo a sharp topological phase transition under tuning of the electrical field through the critical field. We will also present the phase diagram of the nanolayer from which a guideline of finding optimal conditions to achieve a large topological bandgap can be extracted.

{\it Band inversion and topological phase transition}.---To demonstrate the principal concept, we consider an InSb nanolayer with a finite thickness but with infinite planar dimensions under application of a perpendicular electric field [as indicated in the schematic of Fig.~\ref{Fig1}(a)]. Thus, the nanolayer can be effectively viewed as a 2D lattice with the size of a unit cell  proportional to the layer thickness. The 2D band structures are calculated under different strengths of the electric field and the topological phase transition are analyzed by examining the electric-field induced variation of the bandgap and the characteristic topological invariant.
The renowned $sp^{3}s^{\ast}$ tight binding (TB) method~\cite{Vogl1983} is adopted in the calculations of the band structures and the normal direction of the layer is assumed to be parallel to the crystallographic [001] direction. The obtained results can be straightforwardly extended to the layers with different normal crystallographic directions.

Expanding in the $sp^{3}s^{\ast}$ basis~\cite{Vogl1983}, the Hamiltonian of the system can be written as,
  \begin{align}
  H_{\alpha\nu,\beta\xi}(\mathbf{k})=\sum_{\mathbf{R}}e^{-i\mathbf{k\cdot(\mathbf{R}_{\nu}-\mathbf{R}_{\xi})}}\langle\alpha,\mathbf{R}_{\nu}|H|\beta,\mathbf{R}_{\xi}\rangle,
  \label{H-1}
  \end{align}
where $|\alpha,\mathbf{R}_{\nu(\xi)}\rangle$ represents  spin-orbital basis state $\alpha$ at $\mathbf{R}_{\nu (\xi)}=\mathbf{R}+\mathbf{r}_{\nu (\xi)}$, $\mathbf{R}$ is a lattice vector, $\mathbf{r}_{\nu(\xi)}$ is the displacement of atom $\nu(\xi)$ in a unit cell,  and $\mathbf{k}=(k_{x},k_{y})$ is an in-plane wave vector with $k_x$ and $k_y$ being defined along the crystallographic [1-10] and [110] directions.
The diagonal terms in Eq.~(\ref{H-1}) represent  the on-site atomic orbital energies, while the off-diagonal terms describe the interactions between different intra-atomic orbitals, i.e., the terms arising from the intrinsic SOCs between the intra-atomic $p$ orbitals, and the interactions between nearest-neighbor atomic orbitals~\cite{Persson2006,Liao2016,SM}. The effect of the external electric field is taken into consideration by adding a location-dependent electric potential to the on-site energies~\cite{Xiao2019},
  \begin{align}
 \delta V_{\tau}=e F (\tau-\chi_{0}) a_{l},
  \end{align}
where  $e$ is the elementary charge, $F$ represents the electric field strength, $\tau=1,..., 2N_{ }$ the indices of the atomic layers, $\chi_{0}=N_{ }+1/2 $  an offset constant, and  $a_{l}=a_{0}/4$ the inter-atomic layer distance with $a_0=6.48{\AA}$ denoting the material lattice constant and $h=N_{ }a_{0}/2$ giving the thickness of the nanolayer [see Fig.~\ref{Fig1}(a)]. In order to avoid the effect of surface relaxation and reconstruction on the calculations, the dangling bonds of the surface atoms have been passivated by hydrogen-like  atoms~\cite{Persson2006,Liao2016,SM}.  For further details and the $sp^{3}s^{\ast}$ TB parameters employed in the calculations for the band structures of InSb nanolayers with hydrogen passivated surfaces, we refer to Ref.~\onlinecite{SM}.

\begin{figure}
\includegraphics[width=0.48\textwidth]{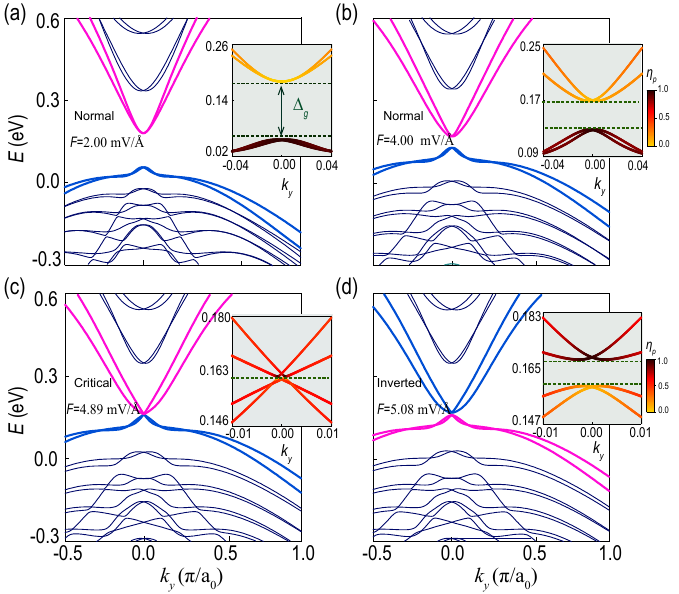}
\caption{(a)-(d) Energy spectra of a [001]  InSb nanolayer containing $2N^{}_{}=60$ atomic layers plotted against $ k^{}_{y}$ at $k^{}_{x}=0$ at different  strengths $F$ of the perpendicularly applied electric field.  Insets show the zoom-in views of  the Bloch  bands near the bandgap. Colors coded to the Bloch bands mark the percentages $\eta^{}_{p}$ of the contributions  from  $p$-orbitals to the Bloch  bands.  }
\label{Fig2}
\end{figure}

To illustrate the bandgap variation with the external perpendicular electric field and the topological phase transition in a semiconductor nanolayer, we show in Fig.~\ref{Fig2} the calculated band structures of the InSb nanolayer with   $2N=60$   at different electric field strengths $F$.
Figure~\ref{Fig2}(a) shows the energy band structure at a small electric field, where $\Delta_{g}$ is the bandgap at the $\Gamma$ point. As depicted in Figs.~\ref{Fig2}(b) and \ref{Fig2}(c), the bandgap $\Delta_{g}$ is reduced with increasing $F$ and gets closed when the field reaches to a {\it critical} value of $F_{\rm cr}\approx4.89$~mV/{\AA}. With further increasing $F$, the conduction and valence bands will be {\it inverted} and the electron-hole hybridization promotes a bandgap reopening, as demonstrated in Fig.~\ref{Fig2}(d). The band inversion can also be manifested by changes in the orbital characters of the Bloch bands near the bandgap \cite{SM}. When $F$ goes beyond the critical field, the bands below and above the bandgap contain contributions dominantly from  the $s$- and $p$-like orbitals, respectively, which is opposite to that in the normal, no-band-inversion cases [see the insets of Fig.~\ref{Fig2}].

For  2D Bloch bands with time-reversal symmetry, the topology is characterized by a single $\mathbb{Z}_{2}$ invariant~\cite{Hasan2010,Qi2011}, with two possibly taking values of 0 and 1, and the nonzero invariant corresponds to the nontrivial phase. The determination of the $\mathbb{Z}_{2}$ invariant involves all the occupied Bloch bands over the half Brillouin zone (BZ)~\cite{Fu2006a}, and the breaking of the inversion symmetry induced by the perpendicular electric field complicates the calculations for the values of the $\mathbb{Z}_{2}$ invariant. However, by regarding one component of the 2D wave vector $\mathbf{k}$ as an virtual periodic time parameter, the invariant can be ascertained by tracking the evolutions of the centers of maximum localized Wannier functions, constructed from the fully occupied bands, over a half period ~\cite{Soluyanov2011,Yu2011,SM}.

\begin{figure}
\includegraphics[width=0.44\textwidth]{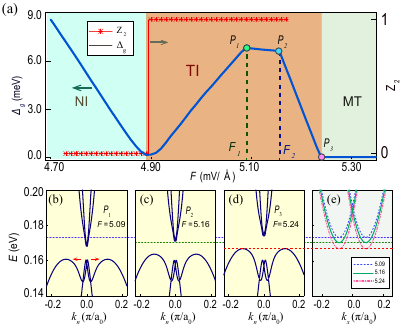}
\caption{(a) Bandgap $\Delta^{}_{\rm g}$ and topological invariant  $\mathbb{Z}_{2}$  plotted against  electric field  $F$ for  the same InSb nanolayer as in  Fig.~\ref{Fig2} with NI labeling the trivial insulating phase, TI the topological insulating phase, and MT the metallic phase.  (b)-(d) Energy spectra of the  inverted Bloch bands  (around the $\Gamma$ point) of the nanolayer  at three  different  electric fields $F_1$, $F_2$ and $F_3$ and the three corresponding sharp inflection points $P_1$, $P_2$ and $P_3$ in the $\Delta^{}_{\rm g}$ curve marked in (a). Here, the bands are plotted against ${\bf k}_n = k_n {\bf n}$, where unit vector ${\bf n}=(2/\sqrt{5}\, {\bf a}_{[1-10]}, 1/\sqrt{5}\, {\bf a}_{[110]},\, 0\,)$ with ${\bf a}_{[1-10]}$ and ${\bf a}_{[110]}$ being the unit vectors along the [1-10] and [110] crystallographic directions. (e) Corresponding energy  spectra of the lowest conduction band (near the $  L  $ point) versus $k^{ }_{x}$ at $k^{}_{y}=\sqrt{2}\pi/ a^{}_{0}$  and electric fields $F_1$, $F_2$, and $F_3$.   }
\label{Fig3}
\end{figure}

Figure~\ref{Fig3}(a) shows the derived topological invariant $\mathbb{Z}_{2}$  and the bandgap $\Delta _{g}$  as a function of the electric field $F$ for the InSb nanolayer with $2N=60$. Here, an abrupt change in $\mathbb{Z}_{2}$ is observed at $F\approx 4.89$ mV/{\AA}. As the electric field is swept through the critical value of $4.89$~mV/{\AA}, $\mathbb{Z}_{2}$ changes from 0 to 1 and is accompanied by the reopening of the bandgap as a result of electron-hole hybridization. Thus, the InSb nanolayer turns to a 2D topological insulator. It is seen that this reopened topological bandgap increases with increasing $F$ but reaches to a \textit{maximum} value of $\Delta_{\max}=6.73$~meV at $F^{}_{\rm 1}\simeq5.09$~mV/{\AA} [i.e., at point $P_1$ in Fig.~\ref{Fig3}(a)]. A further increase of $F$ leads to a decrease in bandgap. This is due to the fact that when $F$ goes beyond  $F^{}_{\rm 1}$, the top of valence bands appears no longer around the $\Gamma$ point, but at finite $k$ points [see Fig.~\ref{Fig3}(b)], and these new top valence states become to interact more weakly with the bottom conduction band states. However, it is found that the topological bandgap  $\Delta^{}_{\rm g}$ is confronted with a quick decrease as $F$ exceeds a value of $F^{}_{\rm 2}\simeq 5.16$~mV/{\AA}. The  physics behind this is that when $F$ reaches $F^{}_{2}$, the conduction band valley at the $L$ point gets aligned with the bottom of the conduction band around the $\Gamma$ point [see Figs.~\ref{Fig3}(c) and \ref{Fig3}(e)]. These conduction band $L$-valley states have a much heavier effective electron mass~\cite{Klimeck2000} and move towards low energies with increasing $F$. Eventually, the topological band-inverted energy gap gets \textit{closed} when $F$ reaches a value of $5.24$mV/{\AA}, at which the $L$-valley conduction band states get aligned with the top of the valence bands in energy [see Figs.~\ref{Fig3}(d) and \ref{Fig3}(e)], and the InSb nanolayer turns to be at a metallic state when $F$ goes beyond $5.24$mV/{\AA}.

\begin{figure}
\includegraphics[width=0.48\textwidth]{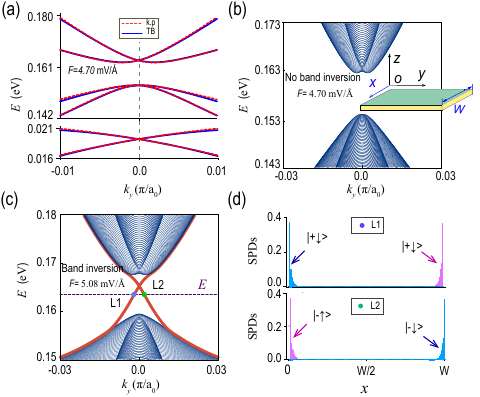}
\caption{(a) Energy bands (solid lines) of the InSb nanolayer as in Fig.~\ref{Fig2} obtained by the tight-binding calculations at $F=4.70$~mV/{\AA} and corresponding band curves (dashed lines) obtained based on the fitted \textbf{k$\cdot$p} Hamiltonian of Eq.~(\ref{H-E}). (b) Energy subbands of an  InSb  nanolayer Hall bar with a transverse width of $W=1.4$~$\mu$m (as shown in the inset) in a topologically trivial phase with electric field $F=4.70$~mV/{\AA}. (c) The same as in (b), but with an inverted  bandgap under a larger field of $F=5.08$~mV/{\AA}. (d) Spatial probability  distributions (SPDs)  of the wave functions of the four gapless band states observed in (c) at a fixed energy of $E^{}_{}=0.163$eV [as  indicated by points $L1$ and $L2$ in (c)]. Here,  $|\pm\uparrow\rangle$ ($|\pm \downarrow\rangle$) label  the edge  states  with the spin  closely parallel (antiparallel) to the $z$-axis and  propagating  along the positive and negative $y$-directions.      }
\label{Fig4}
\end{figure}

{\it Edge states in an InSb nanolayer with a finite width.}---The nontrivial topology can also be manifested by the presence of edge states and the quantum spin Hall effect in a band-inverted InSb nanolayer Hall-bar structure. To demonstrate this, we consider an InSb nanolayer with a finite width as shown in the inset of Fig.~\ref{Fig4}(b). In principle, the energy spectrum of the system can be obtained by the TB method. However, due to lack of translational symmetry in the planar transverse direction and thus an explosively enlarged unit cell, it requires a significantly large amount of computing resources to obtain the band spectrum. Here, to overcome this difficulty, we employe an effective \textbf{k$\cdot$p} Hamiltonian, which is obtained by fitting to the inverted  bands of a corresponding 2D InSb nanolayer obtained by the full TB calculations.

Since only the topologically nontrivial edge states in the band-inverted energy gap will be considered, the effective  Hamiltonian  can be represented in the six-band basis  of  $ \{|\Gamma_{6},1/2\rangle,|\Gamma_{6},-1/2\rangle,|\Gamma_{8},3/2\rangle,|\Gamma_{8},1/2\rangle,
|\Gamma_{8},-1/2\rangle,|\Gamma_{8},-3/2\rangle  \} $. Up to second order in the in-plane wave vector $\mathbf{k}$, the Hamiltonian   can be written as,
\begin{align}
H^{}_{\rm eff}(\mathbf{k})=\left(\begin{array}{cc}
H^{}_{6c6c}~ &~H^{}_{6c8v} \\
H^{\dagger}_{6c8v}~ & ~H^{}_{8v8v}
\end{array}\right) \ ,
\label{H-E}
\end{align}
with $H^{}_{6c6c}(\mathbf{k})=\varepsilon^{}_{0}(k) +[(\alpha^{}_{e}k^{}_{-}+\beta^{}_{e}k^{}_{+})\sigma^{}_{+}+\rm{ h.c.}]$, $H^{}_{8v8v}(\mathbf{k})=\varepsilon^{}_{1}(k) +\varepsilon^{}_{2}(k) (J^{2}_{z}-5/4)+\big[\zeta^{}_{1}k^{}_{+}\{J^{}_{z},J^{}_{-}\}+\zeta^{}_{2}
k^{2}_{+}
J^{2}_{-}+\alpha^{}_{h}J^{3}_{+}k^{}_{-}+\beta^{}_{h }(J^{}_{+}J^{}_{-}J^{}_{+}-3J_{+})k^{}_{-}+{\rm h.c.} \big]$, and $H^{}_{6c8v}(\mathbf{k})=\sqrt{3}P (k^{}_{x}T^{}_{x}+k^{}_{y}T^{}_{y})$. Here, $k^{2} =k^{2}_{x}+k^{2}_{y}$, $k_{\pm}=k_{x}\pm ik_{y}$, $\varepsilon^{}_{i=0,1,2}(k)= \varepsilon^{}_{i}+M_{i}k^{2} $,    $\boldsymbol{\sigma^{}_{}}=\{\sigma^{}_{x},\sigma^{}_{y},\sigma^{}_{z}\}$  are the Pauli matrices in   the  electron basis  and $\sigma_{\pm}=\sigma^{}_{x}\pm i\sigma^{}_{y}$, $\boldsymbol{J}=\{J^{}_{x},J^{}_{y},J^{}_{z}\}$   the angular momentum matrices in the $J=3/2$ angular momentum hole basis and $J^{}_{\pm}=J^{}_{x}\pm iJ^{}_{y}$, $\alpha^{}_{e}$ ($\alpha^{}_{h}$) and $\beta^{}_{e}$ ($\beta^{}_{h }$) are  the strengths  of the electron (hole) Rashba and  Dresselhuas SOCs,  $\zeta^{}_{1,2}$  denote the  strengths of other coupling interactions within the  hole basis,  $T^{}_{x, y}$ correspond to  the $2\times 4$  matrices describing the hybridization of the electron  and hole basis states, and $\sqrt{3}P$ denote the hybridization strength~\cite{Winkler2003}. Note that this effective Hamiltonian $H^{}_{\rm eff}(\mathbf{k})$ is reduced to the Bernevig-Hughes-Zhang (BHZ) Hamiltonian ~\cite{Bernevig2006}, when projected to a  lower-dimensional  basis  of $\{|\Gamma^{}_{6},1/2\rangle,|\Gamma^{}_{8},3/2\rangle,|\Gamma^{}_{6},-1/2\rangle,|\Gamma^{}_{8},-3/2\rangle\}$.

Under different electric fields, the coefficients in Eq.~(\ref{H-E}) can be ascertained by fitting the energy eigenvalues of $H_{\rm eff}(\mathbf{k})$ to the energy band curves obtained by the TB method, as seen in Fig.~\ref{Fig4}(a). For the specific values of the fitting parameters of the \textbf{k$\cdot$p} Hamiltonian under different electric fields, we refer Ref.~\onlinecite{SM}. Concretely, for an InSb nanolayer Hall bar with a finite width, the energy in the transverse direction is quantized and the energy spectrum appears in the form of one-dimensional subbands. Figures~\ref{Fig4}(b) and \ref{Fig4}(c) show the subband structures obtained using the effective Hamiltonian $H_{\rm eff}(\mathbf{k})$ for an InSb nanolayer with a finite width under applied electric fields without and with induced band inversion, respectively. It is evidently seen in Fig.~\ref{Fig4}(c) that being different from the case without the band inversion,  there exist four gapless energy bands inside the band-inverted energy gap. These four gapless energy bands exhibit linear energy dispersion relations (the Dirac bands) near the crossing point (the Dirac point). Figure~\ref{Fig4}(d) shows the probability distributions of the four gapless band states at a fixed energy $E$ near the Dirac point.  Due to the  hybridization between the  electron and hole basis states, we can introduce  a composite vector operator ${\cal S}=(\boldsymbol{\sigma}/2)\oplus \boldsymbol{J}$ to calibrate the effective spin direction of the gapless states \cite{SM}. In the band-inverted nanolayer, it is confirmed that the  two  states residing at each side  of the Hall bar are characterized by different spin orientations (which are closely parallel and antiparallel to the $z$-axis) and distinct propagating directions, i.e., by the quantum spin Hall edge states.

\begin{figure}
\includegraphics[width=0.42\textwidth]{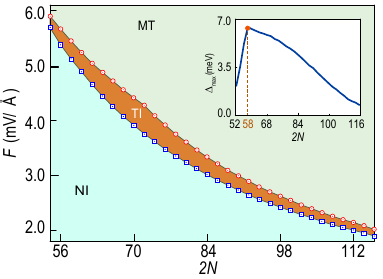}
\caption{Phase diagram of a [001] InSb nanolayer as a function of layer thickness $2N$ and perpendicular electric-field  strength $F$. Here the region for the topological insulator  phase (TI) is highlighted by  dark color. The open symbols are the calculated values for the boundaries of the TI phase and the lines are the guides to the eyes. The regions for the trivial insulating  phase (NI) and the metallic phase (MT) are located below and above the TI phase region. The inset shows the maximum value of the topological gap $\Delta^{}_{\rm max}$ as a function of layer thickness $2N$.}
\label{Fig5}
\end{figure}

{\it Phase diagram of an InSb  nanolayer.}---Figure~\ref{Fig5} displays the phase diagram of the InSb nanolayer with changing nanolayer thickness and electric field. It is shown that the  strength of the critical field at which the phase transition occurs decreases with increasing layer thickness. It is also clearly seen that there presents a stripe of the topologically nontrivial phase (dark colored TI region) in the phase diagram, separating the regions of the trivial insulator phase (NI region) and metallic phase (MT region). The inset of Fig.~\ref{Fig5} shows the maximum value of the topological gap $\Delta_{\rm max}$, which could be found in the topological phase, as a function of layer thickness. It is found that $\Delta_{\rm max}$ increases with increasing layer thickness for a thin InSb nanolayer and, after reaching a value of $\sim 6.8$~meV, $\Delta_{\rm max}$ turns to decrease with increasing layer thickness. In principle, the size of the topological gap $\Delta_{\rm max}$ is largely related to the hybridization of the electron and hole states located close to the two opposite surfaces after the band inversion. Thus, $\Delta_{\rm max}$ would continuously increase with shrinking layer thickness. However, due to the presence of the $L$-valley conduction band states with a heavy effective mass and the fact that the upper inverted bands near the $\Gamma$ point has been largely pushed up by quantum confinement, the size of the topological gap has been compromised by the conduction band states near the $L$-valley [cf. Figs.~\ref{Fig3}(d) and \ref{Fig3}(e)]. Similarly, for an InSb nanolayer with a sufficiently small layer thickness, the stripe width of the topological insulator phase is not increased with decreasing layer thickness as one would expect, but instead it is decreased with decreasing layer thickness due to the presence of the low-energy, heavy $L$-valley electron states.

{\it Discussion and conclusion}.---In this work, we have shown that the topologically nontrivial phase can be established in a conventional, narrow bandgap semiconductor nanolayer through application of a perpendicular electric field. The topological phase transition occurs in such a nanolayer when the applied electric field goes beyond a critical strength at which the conduction and valence bands of the nanolayer are inverted.  In order to obtain a substantially large  inverted  topological bandgap,  it is preferable to explore a semiconductor nanolayer with a small thickness, as to  amplify  the degree of the electron-hole hybridization. However, in this case, a combined effect of strong quantum confinement and requirement for applying a large field may result in an unfavorable situation, with a conduction band valley lying below the inverted upper-band minimum, and thus prevent from achieving a desired large topological bandgap. Thus, there is a trade-off between the size of the  topological bandgap  and the size of the topological phase region (cf.~the phase diagram in Fig.~\ref{Fig5}).
Overall,  our finding   should promote the exploiting of the  topological phase transition in conventional semiconductor nanolayers for novel physics studies and for topological quantum computing technology developments. It is also possible to extend the  expedient  proposal  to other layered materials with a suitable-size bandgap.

\begin{acknowledgments}

\textit{Acknowledgments.}---We thank P.~Kotetes, D.~E.~Liu,  Q.~F.~Sun and J.~H.~Zhao for useful discussions. This work is supported by the National Natural Science Foundation of China (Grant Nos.~92165208, 11874071, 11974340 and 92265203), the Key-Area Research and Development Program of Guangdong Province (Grant No.~2020B0303060001), and the Chinese Academy of Sciences (Grant Nos.~XDB28000000,  XDPB22 and QYZDJ-SSWSYS001).

\end{acknowledgments}

\end{document}